\documentclass[10pt,a4paper]{article}
\usepackage{graphicx}
\date{}
\begin{document}

\date{}

\title{The Dark Mass Problem\\
{\normalsize Solved ?}}
\author{Ll. Bel\thanks{e-mail:  wtpbedil@lg.ehu.es}}

\maketitle

\begin{abstract}

I discuss some of the basic properties of a potential theory derived from a modified Newton's law of action at a distance that includes a $1/r$ attractive force.
\end{abstract}

\section{Point particles}

Let us start assuming that two point particles with masses $m_1$ and $m_2$ located at positions $x^i$ and $y^i$ attract each other with a force $F^i$:

\begin{equation}
\label{Force}
F_{y|}^i(x)=-\frac{G m_1m_2}{r^3}(x^i-y^i)-\frac{G^\prime m_1m_2}{r^2}(x^i-y^i), \quad i,j,\cdots=1,2,3
\end{equation}
where $r^2=|\overrightarrow{x}-\overrightarrow{y}|^2$, $G$ is Newton's constant and $G^\prime$ is a free positive physical constant with dimensions M${}^{-1}$L${}^{2}$T${}^{-2}$.

This law of attraction has been considered, more or less directly, as a candidate to solve what is known today as the Dark mass problem (more on that below).

In considering this new law of force it is important to keep in mind that space has three dimensions and not two, and as a consequence of this  only the Newtonian component of (\ref{Force}) is solenoidal.

Since:
\begin{equation}
\label{Curl}
\frac{\partial  F_{y|i}(x)}{\partial x^j}-\frac{\partial  F_{y|j}(x)}{\partial x^i}=0,
\end{equation}
introducing the potential function $V$ defined by:

\begin{equation}
\label{Potential}
F_{y|i}(x)=-m_1\frac{\partial  V_{y|}(x)}{\partial x^i}
\end{equation}
from (\ref{Force}) there follows  that:

\begin{equation}
\label{Laplacian}
\Delta  V_{y|}(x)=4\pi G\rho_{y|}(x),  \quad \rho_{y|}(x)\equiv m_2\delta(r)+\frac{\alpha m_2}{r^2}, \quad \alpha=\frac{1}{4\pi}\frac{G^\prime}{G}
\end{equation}

Let us assume that the point particle with mass $m_2$ is kept fixed at the location $y^i$ and that the particle $m_1$ is free to move under its attraction. This simple formula above tells us that this particle will move without friction as it would do across a cloud of dark matter with an always positive effective density $\rho$, while obeying a potential theory formally identical to Newton's one.

\section{Extended sources}

If instead of a point particle with mass $m_2$ we consider a continuous distribution of mass with density $\mu(y)$ then we shall have:

\begin{equation}
\label{Mu}
 F^i(x)=-G m_1\int_D \mu(y)\frac{(x^i-y^i)}{r^3}\,d^3y-G^\prime m_1\int_D \mu(y)\frac{(x^i-y^i)}{r^2}\,d^3y,
\end{equation}
where $D$ is the domain where $\mu\neq 0$, and:

\begin{equation}
\label{Mu}
V(x)=-G\int_D \mu(y)\frac{1}{r}\,d^3y+G^\prime\int_D \mu(y)\ln(r)\,d^3y,
\end{equation}
from where we get:

\begin{equation}
\label{Extended}
\Delta  V(x)=4\pi G\rho(x),  \quad \rho(x)\equiv \mu(x)+\alpha\int_D \frac{\mu(y)}{r^2})\,d^3y,
\end{equation}

Let us consider in particular the case where the density $\mu$ is constant inside a sphere of radius $a$ and zero otherwise (Figure 1). In this case, $r$ being now $|\overrightarrow{x}|$, the effective density $\rho$ will be:

\begin{equation}
\label{Integral}
\rho(r)=\mu H(a-r)+\alpha\int_0^{2\pi}d\phi\int_0^a du\int_0^\pi d\theta \frac{\mu u^2\sin \theta}{r^2+u^2-2ru\cos \theta}
\end{equation}
$H$ being the Heaviside function; or:

\begin{equation}
\label{Density}
\rho(r)=\mu H(a-r)+2\pi\alpha\mu\sigma(r,a)
\end{equation}
with:

\begin{equation}
\label{sigma}
\sigma(r,w)=w+\frac{1}{2r}(r^2-w^2)\ln\left(\frac{|r-w|}{r+w}\right)
\end{equation}
from where, to calculate the central potential or the force, we could proceed as usual integrating the Poisson's equation (\ref{Extended}), or use their  corresponding integral definitions.

If instead we have an spherical sector with inner radius $b$ (Figure 2) then the effective density $\rho(r)$ is:

\begin{equation}
\label{Sector}
\rho(r)=\mu H(a-r)H(r-b)+2\pi\alpha\mu(\sigma(r,a)-\sigma(r,b))
\end{equation}

Notice that in the cavity defined by $r<b$, where the raw matter density is zero, the effective density remains positive and therefore the force remains attractive towards the center. This means that it is a legitimate speculation to point out that if gravity includes a contribution of the type that I have considered here, then we could sometimes be fooled  to believe that a massive object gravitates around a source that actually does not exists at all.

\section{Comments}

Notable consequences of the  proposed new law are:

i) If $G^\prime\neq 0 $ and there is ordinary matter somewhere then there is Dark matter everywhere. Therefore this new law has a lot to say about the rotation curves of spiral galaxies as well as micro and macro lensing. The price that we pay for it is the non locality of the theory since the effective density is not a point function.

ii) It predicts new surprising effects like massive bodies orbiting central un-existing matter sources in empty cavities (Empty cavity effect), and beyond that it promises an interesting development of the physics of voids in cosmology..

iii) Last but not least, the transport of this proposal to General relativity is quite natural. It suffices to use the effective density $\rho(x)$ instead of the pure matter one $\mu$ in the source energy-momentum tensor of Einstein's  equations. The first conceptually important implications are: 1) that the gravitational field of a point particle becomes more singular than what it is in Einstein's theory and 2) that, so to speak, all exact vacuum solutions become approximate approximate solutions of the modified theory. On the other hand cosmology only needs to sort out what part of the total density is matter density and what part it is dark matter.

Reference \cite{Kinney} considers a theory based from the beginning on the potential:

\begin{equation}
\label{TotalForce}
V(r)=-\frac{G m_1m_2}{r}+G^\prime \ln r
\end{equation}
Some aspects of this point of point view bear a similitude with mine but it is by no means equivalent to it. It does not predict any Empty cavity effect.

References \cite{Kirillov1} and \cite{Kirillov2} are based on a Modified quantum field theory that leads them to describe the gravity of point particles by a potential:

\begin{equation}
\label{Kirillov}
V(r)=\frac{1}{2\pi^2}\int_0^\infty(V(\omega)\omega^3)\frac{\sin(\omega r)}{\omega r}\frac{d\omega}{\omega}.
\end{equation}
They claim that this potential can be approximated by a $1/r^2$  or a $1/r$ one depending on the scale of distances. They also note "that from a dynamical point of view  the modification of the Newton's law of gravity can be interpreted as if point sources lose their point-like character and acquire an additional distribution in space". This is also the main point in my very much simpler proposition.

Reference \cite{Fabris} starts with the introduction of the potential function (\ref{Mu}) but the only source model that is considered is that of a thin disk of matter  that can be dealt with in the framework of potential theory in dimension 2.

\vspace{1cm}

\hspace{-3cm}\includegraphics[scale=.5,angle=0]{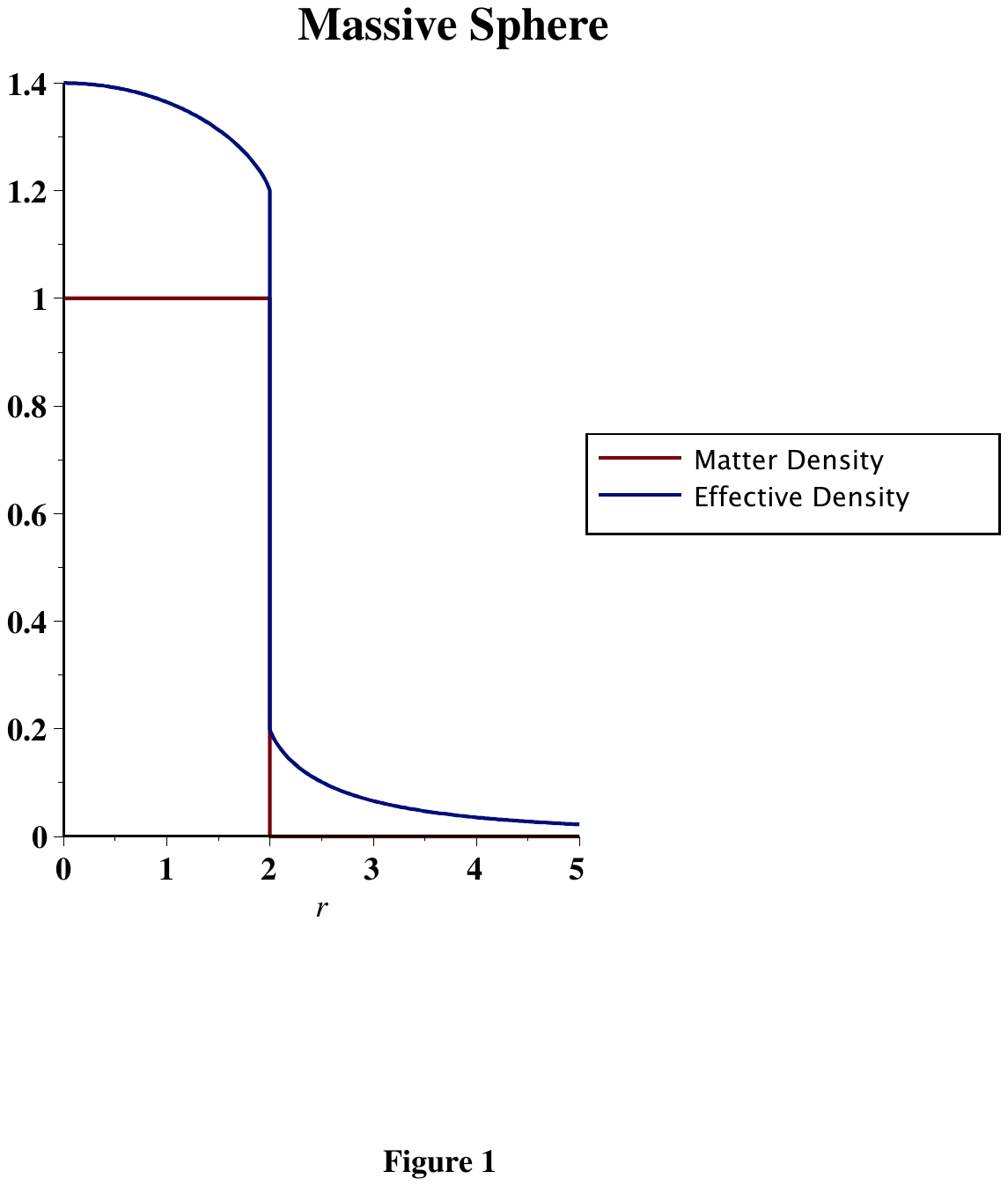}
\hspace{-3cm}\includegraphics[scale=.5,angle=0]{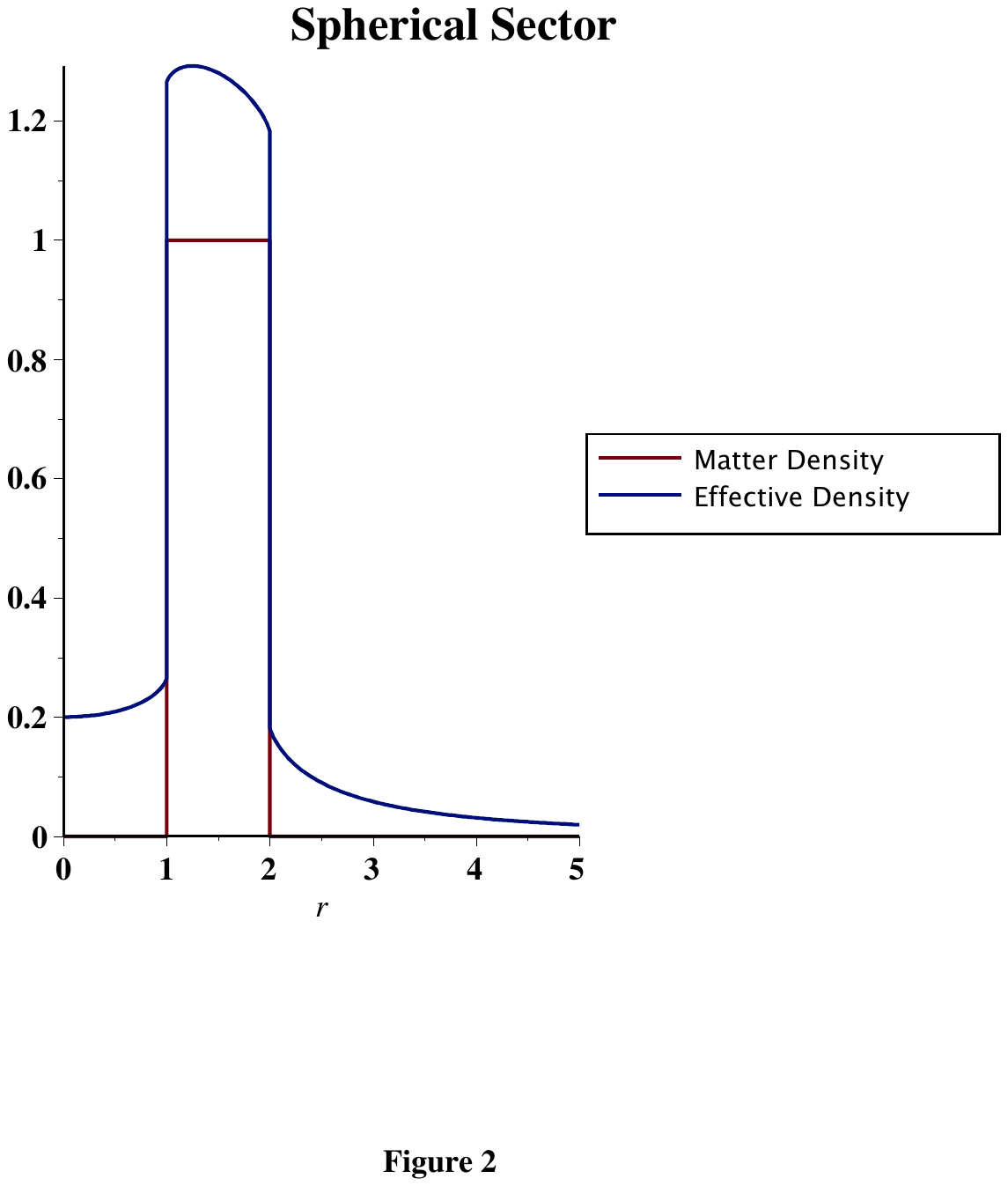}

\end{document}